\documentclass[8.5pt,twoside,twocolumn,amsmath,amssymb]{article}
\usepackage[super,sort&compress,comma]{natbib}
\usepackage{mhchem}
\usepackage{times,mathptmx}
\usepackage{sectsty}
\usepackage{balance}

\usepackage{graphicx} 
\usepackage{lastpage}
\usepackage[format=plain,singlelinecheck=false,font=small,labelfont=bf,labelsep=space]{caption}
\usepackage{fancyhdr}
\usepackage{dcolumn}
\usepackage{bm}
\usepackage{amsmath}
\usepackage{bbding}
\usepackage{verbatim}

\newcommand{\etalia}{{\it et al.}}
\newcommand{\la}{\left\langle}
\newcommand{\ra}{\right\rangle}

\newcommand{\PRL}{Phys.~Rev.~Lett.}
\newcommand{\PR}{Phys.~Rev.~}
\newcommand{\JCP}{J.~Chem.~Phys.}

\newcommand{\JPCM}{J.~Phys.: Condens.~Matter}

\begin{document}

\thispagestyle{plain}
\fancypagestyle{plain}{
\fancyhead[L]{\includegraphics[height=8pt]{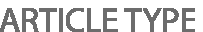}}
\fancyhead[C]{\hspace{-1cm}\includegraphics[height=20pt]{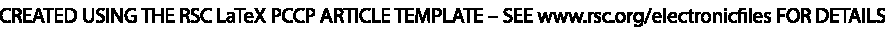}}
\fancyhead[R]{\includegraphics[height=10pt]{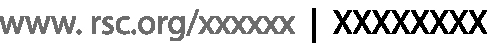}\vspace{-0.2cm}}
\renewcommand{\headrulewidth}{1pt}}
\renewcommand{\thefootnote}{\fnsymbol{footnote}}
\renewcommand\footnoterule{\vspace*{1pt}%
\hrule width 3.4in height 0.4pt \vspace*{5pt}}
\setcounter{secnumdepth}{5}

\makeatletter
\def\subsubsection{\@startsection{subsubsection}{3}{10pt}{-1.25ex plus -1ex minus -.1ex}{0ex plus 0ex}{\normalsize\bf}}
\def\paragraph{\@startsection{paragraph}{4}{10pt}{-1.25ex plus -1ex minus -.1ex}{0ex plus 0ex}{\normalsize\textit}}
\renewcommand\@biblabel[1]{#1}
\renewcommand\@makefntext[1]%
{\noindent\makebox[0pt][r]{\@thefnmark\,}#1}
\makeatother
\renewcommand{\figurename}{\small{Fig.}~}
\sectionfont{\large}
\subsectionfont{\normalsize}

\fancyfoot{}
\fancyfoot[LO,RE]{\vspace{-7pt}\includegraphics[height=9pt]{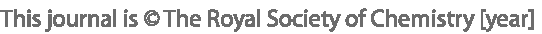}}
\fancyfoot[CO]{\vspace{-7.2pt}\hspace{12.2cm}\includegraphics{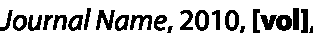}}
\fancyfoot[CE]{\vspace{-7.5pt}\hspace{-13.5cm}\includegraphics{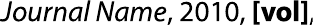}}
\fancyfoot[RO]{\footnotesize{\sffamily{1--\pageref{LastPage} ~\textbar  \hspace{2pt}\thepage}}}
\fancyfoot[LE]{\footnotesize{\sffamily{\thepage~\textbar\hspace{3.45cm} 1--\pageref{LastPage}}}}
\fancyhead{}
\renewcommand{\headrulewidth}{1pt}
\renewcommand{\footrulewidth}{1pt}
\setlength{\arrayrulewidth}{1pt}
\setlength{\columnsep}{6.5mm}
\setlength\bibsep{1pt}

\twocolumn[
  \begin{@twocolumnfalse}
\noindent\LARGE{\textbf{Influence of Polymer Shape on Depletion Potentials and Crowding
in Colloid-Polymer Mixtures}}
\vspace{0.6cm}

\noindent\large{\textbf{Wei Kang Lim\textit{$^{a}$} and Alan R. Denton\textit{$^{\ast}$\textit{$^{a}$}}
}}\vspace{0.5cm}

\noindent\textit{\small{\textbf{Received 23rd November 2015, Accepted 14th December 2015\newline
First published on the web 15th December 2015}}}

\noindent \textbf{\small{DOI: 10.1039/b000000x}}
\vspace{0.6cm}

\noindent \normalsize{
Depletion-induced interactions between colloids in colloid-polymer mixtures 
depend in range and strength on size, shape, and concentration of depletants. 
Crowding by colloids in turn affects shapes of polymer coils, such as biopolymers 
in biological cells.  By simulating hard-sphere colloids and random-walk polymers, 
modeled as fluctuating ellipsoids, we compute depletion-induced potentials and 
polymer shape distributions.  Comparing results with exact density-functional theory calculations, 
molecular simulations, and experiments, we show that polymer shape fluctuations 
play an important role in depletion and crowding phenomena.
}
\vspace{0.2cm}
 \end{@twocolumnfalse}
  ]

\footnotetext{\textit{$^{a}$~Department of Physics, North Dakota State University, 
Fargo, ND 58108-6050, USA.  E-mail: alan.denton@ndsu.edu}}

\section{Introduction}

Depletion forces are ubiquitous in soft materials that contain hard particles and 
flexible macromolecules,\cite{lekkerkerker-tuinier2011}
such as colloid-polymer and colloid-surfactant mixtures.
Over 60 years ago, Asakura and Oosawa\cite{asakura1954} recognized that
the exclusion of one species (depletant) from the space between two particles of another 
species creates an osmotic pressure imbalance that induces an entropy-driven 
attraction between the particles and can drive demixing into
colloid-rich and colloid-poor phases.\cite{vrij1976,pusey1991}
Practical applications of depletion forces are in initiating flocculation of impurities 
in water treatment and winemaking, promoting aggregation of DNA and crystallization 
of proteins,\cite{zukoski2001}
and controlling stability and dynamical properties of many consumer products,
including paints, foods, and pharmaceuticals.\cite{lekkerkerker-tuinier2011}
Depletion forces have been measured by several experimental methods,
including total internal reflection microscopy,\cite{leiderer1998}
atomic force microscopy,\cite{milling-biggs1995}
neutron scattering,\cite{tong1996} and optical trapping.\cite{yodh1998,yodh2001,dogic2015}
Modeling efforts have invoked force-balance theory,\cite{walz-sharma1994,piech-walz2000}
perturbation theory,\cite{mao-cates-lekkerkerker1995,mao-cates-lekkerkerker1997}
polymer field theory,\cite{eisenriegler1996,hanke1999,odijk2000,fleer-tuinier2003,
eisenriegler2003,forsman2014} density-functional theory,\cite{leiderer1999,forsman2009}
adsorption theory,\cite{tuinier-lekkerkerker2000,tuinier-petukhov2002}
integral-equation theory,\cite{chatterjee1998a,schweizer2002,moncho-jorda2003}
Monte Carlo simulation methods,\cite{meijer-frenkel1994,meijer-frenkel1991,dickman1994,
louis2000-prl,bolhuis2001-jcp,bolhuis2001-pre,bolhuis-louis-hansen-prl2002,louis2002-jcp1,
bolhuis-louis2002-macromol,louis2002-jcp2,bolhuis2003, doxastakis2005,likos2010} and 
free-volume theories for thermodynamic phase behavior.\cite{lekkerkerker1992,aarts2002,fleer-tuinier2008}

Complementary to depletion is the phenomenon of crowding upon mixing polymers or other 
flexible macromolecules with impenetrable obstacles.  When colloids, nanoparticles, 
or other crowding agents are dispersed in a polymer solution or blend, flexible 
chains adjust their size and shape to conform to the accessible volume.\cite{denton-cmb2013}
The prevalence and importance of macromolecular crowding in biology 
was recognized over three decades ago.\cite{minton1981}  In the congested 
environment of a cell's cytoplasm or nucleoplasm, conformations of proteins, 
RNA, and DNA are constrained by the presence of other macromolecules, affecting 
biopolymer function.\cite{ellis2001b,cheung2013,denesyuk-thirumalai2013a}
Crowding of polymers has been studied experimentally by 
neutron scattering,\cite{kramer2005a,longeville2009}
computationally via Langevin dynamics\cite{cheung2005,wittung-stafshede2012,
denesyuk-thirumalai2013b} and 
Monte Carlo simulations of coarse-grained models,\cite{hoppe2011,lu-denton2011,lim-denton2014}
and by free-volume theories.\cite{denton-cmb2013,minton2005,lu-denton2011,lim-denton2014}

Attempts to interpret experimental or simulation data for depletion forces in colloid-polymer 
mixtures typically assume the spherical polymer model and treat the polymer size and concentration 
as free parameters.\cite{leiderer1998,milling-biggs1995,tong1996,yodh1998,yodh2001}
The fitted parameters invariably differ from measured values.
Dependences on particle curvature and depletant concentration have been partially 
accounted for by introducing an effective polymer size or depletion layer 
thickness.\cite{louis2002-jcp1,louis2002-jcp2,aarts2002,fleer-tuinier2003}
The influence of depletant shape on interactions has been studied in mixtures 
of colloidal spheres and rods\cite{mao-cates-lekkerkerker1997,yodh2001} 
or ellipsoids,\cite{kamien1999,piech-walz2000} but only of fixed size and shape.
Other workers have explored the impact of polymer conformations on relative stabilities 
of hard-sphere colloidal crystals\cite{panagiotopoulos2014,panagiotopoulos2015-1,panagiotopoulos2015-2}
and of crowding on polymer size\cite{kramer2005a,
longeville2009,minton2005,cheung2005,wittung-stafshede2012,denesyuk-thirumalai2013b,
hoppe2011,lu-denton2011,lim-denton2014} (but not shape).
Despite ample evidence that random-walk polymers exhibit significant
asphericity,\cite{kuhn1934,solc1971,rudnick-gaspari1987,wirtz2000} however,
no studies have yet related shape fluctuations to depletion interactions and crowding.
This paper presents the first consistent analysis of the role of depletant shape 
in mixtures of colloids and nonadsorbing polymers.  By comparing results with exact 
theoretical calculations and with data from molecular simulations and experiments, 
we demonstrate the importance of polymer shape fluctuations for depletion and crowding.

\section{Model}

Our model generalizes the widely-studied Asakura-Oosawa-Vrij (AOV) model of colloid-polymer 
mixtures,\cite{asakura1954,vrij1976} which represents the colloids as hard spheres 
and the polymers as effective spheres of fixed size that are mutually noninteracting,
but impenetrable to the colloids.  The assumption of hard colloid-polymer interactions 
is reasonable for colloids larger than the polymers.
However, the effective-sphere approximation ignores conformational fluctuations of polymer coils.
Here we go beyond previous coarse-grained models of polymer-induced depletion interactions
by representing the polymers as soft ellipsoids that fluctuate in size and shape.

A polymer coil of $N$ segments has size and shape characterized by its gyration tensor, 
${\bf T}=(1/N)\sum_{i=1}^N{\bf r}_i{\bf r}_i$,
where ${\bf r}_i$ is the position vector of segment $i$ relative to the center of mass.
A conformation with gyration tensor eigenvalues $\Lambda_1$, $\Lambda_2$, $\Lambda_3$
has instantaneous radius of gyration $R_p=\sqrt{\Lambda_1+\Lambda_2+\Lambda_3}$.  
The experimentally measurable (root-mean-square) radius of gyration is an ensemble average 
over polymer conformations, $R_g\equiv\la R_p^2\ra^{\scriptstyle 1/2}$.
If the average is defined relative to the polymer's principal-axis frame,
the coordinate axes being labelled to preserve the eigenvalue order
($\Lambda_1>\Lambda_2>\Lambda_3$), then the average gyration tensor describes an 
aspherical object, whose average shape is an elongated (prolate), 
flattened ellipsoid.\cite{kuhn1934,solc1971,rudnick-gaspari1987}
Each eigenvalue is proportional to the square of a
principal radius of the general ellipsoid that best fits the shape of the polymer:
$x^2/\Lambda_1+y^2/\Lambda_2+z^2/\Lambda_3=3$.

Ideal, freely-jointed (random-walk) polymer coils can be modeled as soft Gaussian 
ellipsoids.\cite{murat-kremer1998}  For coils sufficiently long that extensions 
in orthogonal directions are essentially independent, the shape probability distribution 
is well approximated by the factorized form\cite{eurich-maass2001}
\begin{equation}
P(\lambda_1,\lambda_2,\lambda_3) = P_1(\lambda_1)P_2(\lambda_2)P_3(\lambda_3)~,
\label{Eurich-Maass1}
\end{equation}
where $\lambda_i\equiv\Lambda_i/(Nl^2)$ ($i=1,2,3$) for segment length $l$ and 
\begin{equation}
P_i(\lambda_i) = \frac{(a_id_i)^{n_i-1}\lambda_i^{-n_i}}{2K_i}
\exp\left(-\frac{\lambda_i}{a_i}-d_i^2\frac{a_i}{\lambda_i}\right)~,
\label{Eurich-Maass2}
\end{equation}
with parameters 
$K_1=0.094551$, $K_2=0.0144146$, $K_3=0.0052767$, 
$a_1=0.08065$, $a_2=0.01813$, $a_3=0.006031$,
$d_1=1.096$, $d_2=1.998$, $d_3=2.684$, 
$n_1=1/2$, $n_2=5/2$, and $n_3=4$. 
We emphasize that these distributions, which exhibit broad fluctuations in polymer size 
and shape, are derived from random-walk statistics\cite{murat-kremer1998,eurich-maass2001} 
and will be modified in the presence of crowding agents (e.g., colloids).

The deviation of a polymer's average shape from a sphere is quantified
by the asphericity\cite{rudnick-gaspari1987}
\begin{equation}
A=1-3\frac{\la\lambda_1\lambda_2+\lambda_1\lambda_3+\lambda_2\lambda_3\ra}
{\la(\lambda_1+\lambda_2+\lambda_3)^2\ra}~.
\label{asphericity}
\end{equation}
A perfect sphere ($\lambda_1=\lambda_2=\lambda_3$) has $A=0$, while
an object that is greatly elongated along one axis has $A\simeq 1$.
A mixture of spherical colloids of radius $R_c$ and polymers of uncrowded 
rms radius of gyration $R_g$ is characterized by the number densities,
$n_c$ and $n_p$, and size ratio, $q\equiv R_g/R_c$, of the two species.

\section{Methods}

To explore the influence of polymer shape on depletion-induced interactions between 
colloids, and of crowding on polymer conformations, we developed a Monte Carlo (MC) 
algorithm for simulating mixtures of hard colloidal spheres and ideal polymers, 
whose shape distribution follows Eq.~(\ref{Eurich-Maass1}).
At fixed temperature $T$ and volume, trial displacements of colloids and displacements 
and rotations of polymers are accepted with the Metropolis probability\cite{frenkel-smit2001}
$\min\left\{e^{-\beta\Delta U},~1\right\}$,
where $\beta=1/(k_B T)$ and $\Delta U$ is the associated change in potential energy.
Colloid-colloid and colloid-polymer overlaps yield infinite energy and so are always rejected.
To detect intersection of a polymer with a colloid, we implemented an overlap algorithm
that determines the shortest distance between the surfaces of a sphere and 
a general ellipsoid by numerically evaluating the root of a 6th-order polynomial.\cite{heckbert1994}
For trial rotations, we define the orientation of a polymer by a unit vector ${\bf u}$, 
aligned with the long axis of the ellipsoid, and generate a new (trial) direction 
${\bf u}'=({\bf u}+\tau{\bf v})/|{\bf u}+\tau{\bf v}|$,
where ${\bf v}$ is a randomly oriented unit vector and $\tau$ is a tolerance 
determining the magnitude of the rotation.\cite{frenkel-smit2001}
In addition, we perform trial changes in shape of a polymer coil, from 
gyration tensor eigenvalues $\lambda\equiv(\lambda_1,\lambda_2,\lambda_3)$ 
to new eigenvalues $\lambda'\equiv(\lambda_1',\lambda_2',\lambda_3')$. 
Such trial moves, which entail a change in internal free energy of the coil,\cite{doi-edwards1986}
$F_c=-k_BT\ln P_0(\lambda)$, are accepted with probability 
\begin{eqnarray}
{\cal P}(\lambda\to\lambda') &=& 
\min\left\{e^{-\beta(\Delta F_c+\Delta U)},~1\right\} \nonumber\\[1ex]
&=& \min\left\{\frac{P_0(\lambda')}{P_0(\lambda)}
e^{-\beta\Delta U},~1\right\}~,
\label{shape-variation}
\end{eqnarray}
where $P_0(\lambda)$ is the shape distribution in a reservoir of pure polymer 
[Eq.~(\ref{Eurich-Maass1})].  We assume that a coil of a given shape in the system 
has free energy equal to that of an identically shaped coil 
in the reservoir.\cite{denton-schmidt2002}
Through trial changes in eigenvalues, the polymers evolve toward an equilibrium shape 
distribution, constrained by the presence of crowders (colloids).

\begin{figure*}
\begin{minipage}{4.5cm}
\includegraphics[width=0.6\columnwidth]{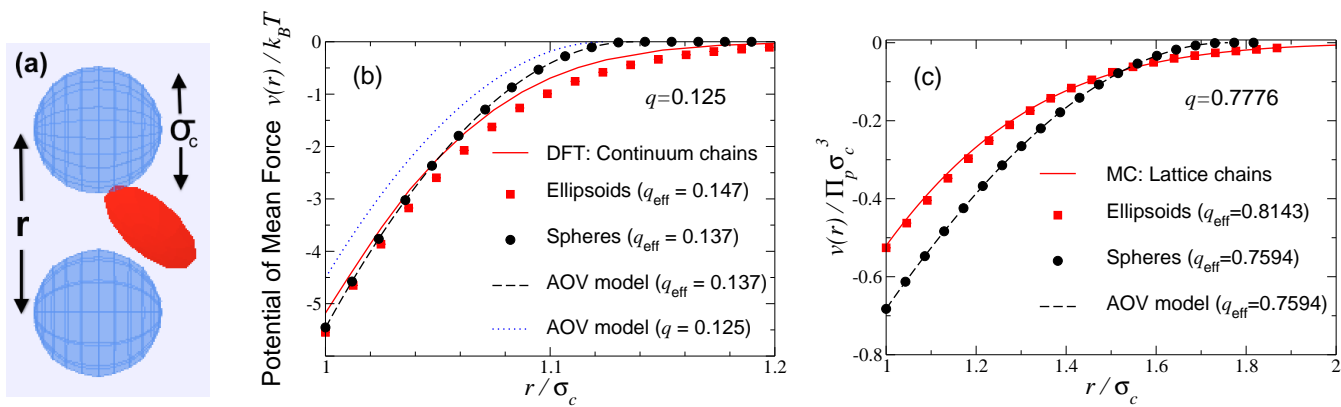} 
\end{minipage}
\hspace*{-1.3cm}
\begin{minipage}{7cm}
\vspace*{-0.32cm}
\includegraphics[width=\columnwidth]{fig1b.eps} 
\end{minipage}
\hspace{0.2cm}
\begin{minipage}{7cm}
\includegraphics[width=\columnwidth]{fig1c.eps} 
\end{minipage}
\caption{
Simulation snapshot (a) depicts colloids (blue spheres) and polymer (red ellipsoid),
which induces potential of mean force $v(r)$ at center-center distance $r$
(units of colloid diameter $\sigma_c$) for polymer-to-colloid size ratios (b) $q=0.125$, (c) $q=0.7776$.
Our MC simulation data for the fluctuating ellipsoidal polymer model (squares), and predictions 
of the spherical polymer (AOV) model (circles, dashed curves), are compared with
(b) density-functional theory (DFT) predictions (solid curve) for continuum-chain 
polymers\cite{forsman2009} and
(c) MC simulation data (solid curve) for lattice-chain polymers\cite{meijer-frenkel1994} 
at corresponding effective size ratios $q_{\rm eff}$ [Eq.~(\ref{qeff})].
Also shown in (b) is the AOV model prediction for the bare size ratio $q$ (dotted curve).
Error bars are smaller than symbols.
\vspace*{0.2cm}
}\label{fig-pmf-frenkel-forsman}
\end{figure*}

Depletion of polymers induces an effective interaction between colloids that reduces, 
in the dilute limit, to the potential of mean force (PMF),
$v_{\rm mf}(r)=\Omega(r)-\Omega(\infty)$, defined as the change in grand potential 
$\Omega(r)$ upon bringing two colloids from infinite to finite
(center-to-center) separation $r$ by working against the polymer osmotic pressure,
$\Pi_p=n_p k_BT$ (for ideal polymers).
If we make the choice $\Omega(\infty)=0$, then $v_{\rm mf}(r)=-\Pi_p V_o(r)$,
where $V_o(r)$ is the intersection of the excluded-volume regions surrounding the colloids.
For spherical polymers (AOV model),
\begin{equation}
V_o(r)=\frac{4\pi}{3}(1+q)^3 R_c^3\left(1-\frac{3r/R_c}{4(1+q)}+
\frac{(r/R_c)^3}{16(1+q)^3}\right)~.
\label{AOV}
\end{equation}
In general, however, since $V_o(r)$ depends on the shapes of colloids {\it and} polymers,
its computation is nontrivial.  In the fluctuating ellipsoidal polymer model, computing 
$V_o(r)$ also requires averaging over polymer shape and orientational distributions.
We determined $v_{\rm mf}(r)$ using a particle insertion method\cite{widom1963,frenkel-smit2001} 
by fixing two colloids at separation $r$ in a simulation box, inserting polymers of 
random shapes and orientations, generated by our MC algorithm, at random positions 
in the space between fixed colloids, and counting the fraction of double overlaps.
Since ideal polymers are independent, we need insert only one at a time and then 
scale by the polymer number.
For the polymer trial moves, we used tolerances of $\tau=0.001$ for rotations and 
$\Delta\lambda_1=0.01$, $\Delta\lambda_2=0.003$, $\Delta\lambda_3=0.001$ for shape changes.
As a check, our algorithm reproduces $V_o(r)$ for spherical polymers (AOV model)
[Eq.~(\ref{AOV})].

Accurate calculation of the PMF requires calibrating the ellipsoidal polymer model to 
consistently match the polymer radius to the depletion layer thickness and to account 
for deformation of a polymer coil near a curved surface.\cite{meijer-frenkel1994}
A rational criterion for choosing the {\it effective} size ratio $q_{\rm eff}$ is 
based on equating the free energy to insert a hard sphere into a bath of ideal polymers, 
as predicted by polymer field theory,\cite{eisenriegler1996} with the work required to 
inflate a sphere in the model polymer solution.\cite{louis2002-jcp1,louis2002-jcp2,aarts2002}
For nonspherical polymers, we generalize this criterion to
\begin{equation}
q_{\rm eff}=\frac{R_g}{c}\left[\left(1+\frac{6}{\sqrt{\pi}}q+3q^2\right)^{1/3}-1\right]~,
\label{qeff}
\end{equation}
where $c$ is the integrated mean curvature of the polymer,\cite{oversteegen-roth2005}
which accounts for shape fluctuations.  We computed $c$ numerically by integrating the 
mean curvature over the ellipsoid surface and averaging with respect to the shape distribution 
[Eq.~(\ref{Eurich-Maass1})], yielding $c=0.93254~R_g$ (compared with $c=R_g$ for spheres 
of fixed radius).  Equation~(\ref{qeff}) ensures that, in the limit $q\to 0$, the model recovers 
the exact depth of the PMF (per unit area) between hard, flat plates 
at contact:\cite{lekkerkerker-tuinier2011,asakura1954} $(4/\sqrt{\pi})R_g\Pi_p$.

\section{Results}

As a first test of the ellipsoidal polymer model, we computed the PMF 
in the dilute colloid limit by performing simulations over a sequence of 
colloid pair separations for the same size ratios as used by 
Forsman and Woodward\cite{forsman2009} in their exact density-functional theory (DFT)
calculations for a continuum-chain polymer model ($q=0.125$) and by
Meijer and Frenkel\cite{meijer-frenkel1994} ($q=0.7776$) in their MC simulations 
of random-walk polymer chains on a cubic lattice.
In each comparison, we used appropriate effective size ratios computed from Eq.~(\ref{qeff})
with $c/R_g=1$ for spheres and $c/R_g=0.93254$ for ellipsoids,
and averaged over five independent runs, each of $2\times 10^7$ polymer insertions.
Figure~\ref{fig-pmf-frenkel-forsman} shows that the PMFs resulting from the 
ellipsoidal polymer model are in excellent agreement with the corresponding PMFs
from both of the explicit polymer models.
Thus, calibration of the effective polymer size near hard, flat plates 
($q=0$) proves accurate also for polymers near hard spheres ($q>0$).
In contrast, the AOV model [Eq.~(\ref{AOV})] predicts a shorter-ranged 
(and, for $q=0.7776$, also deeper) potential, reflecting lack of freedom 
of a spherical polymer to deform to avoid obstacles.

\begin{figure}[t!]
\includegraphics[width=0.9\columnwidth]{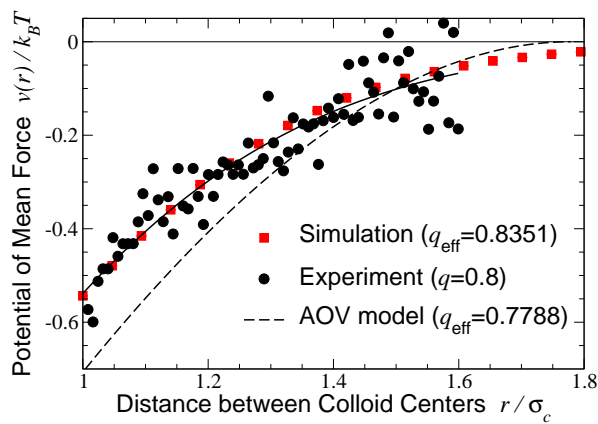} 
\vspace*{-0.1cm}
\caption{
Potential of mean force between a pair of silica microspheres (diameter $\sigma_c=1.25~\mu$m) 
induced by $\lambda$-DNA in water with $R_g=0.5~\mu$m ($q=2R_g/\sigma_c=0.8$).
Our MC simulation data for the fluctuating ellipsoidal polymer model (squares) at
effective size ratio $q_{\rm eff}=0.8351$ [Eq.~(\ref{qeff})] are compared with 
both experimental data\cite{yodh1998} (circles) and predictions of the 
AOV model (dashed curve) [Eq.~(\ref{AOV})] for $q_{\rm eff}=0.7788$ [Eq.~(\ref{qeff})].
The solid curve is a least-squares fit to the experimental ({\it not} simulation) data
of the function $-\exp(a_0+a_1r+a_2r^2)$ with $a_0=0.817$, $a_1=-0.167$, $a_2=-1.269$.
}\label{fig-pmf-dna}
\end{figure}

We turn next to the experiments of Verma \etalia,\cite{yodh1998} who used an
optical tweezer to measure interactions between silica microspheres of diameter 
$\sigma_c=1.25\pm 0.05~\mu$m in aqueous solutions of $\lambda$-DNA (contour length 
16 $\mu$m, radius of gyration $R_g\simeq$ 500 nm), whose conformations are known 
to be random walks of $\sim$160 Kuhn segments.\cite{wirtz2000}
Taking the nominal size ratio of $q=0.8$, we computed the PMF in both the
ellipsoidal and spherical polymer models and here compare our results with data 
for a dilute DNA solution of concentration 25 $\mu$g/ml ($n_p=0.5~\mu$m$^{-3}$), 
in which polymer interactions should be negligible
(Figs.~2 and 3 of ref.~\citenum{yodh1998}).
Since the experiments cannot accurately resolve the vertical offset of the 
potential, we varied the offset to most closely fit our simulation data.  
With this single fit parameter, the ellipsoidal polymer model,
with effective size ratio $q_{\rm eff}=0.8351$ [from Eq.~(\ref{qeff})], 
is in close agreement with the measured interaction potential (Fig.~\ref{fig-pmf-dna}),
as is seen by comparing the least-squares fit to the experimental data 
with our simulation data (solid curve and squares in Fig.~\ref{fig-pmf-dna}).
In contrast, the AOV model, with $q_{\rm eff}=0.7788$, significantly overestimates 
the depth, and underestimates the range, of the potential.  From visual inspection, 
it is clear that no vertical shift of the experimental data will yield close alignment 
with the AOV model (solid and dashed curves in Fig.~\ref{fig-pmf-dna}).

The close agreement of depletion potentials from our simulations of the ellipsoidal 
polymer model with, on the one hand, DFT calculations and simulations for explicit 
polymer models and, on the other hand, experimental data from optical tweezer 
measurements of colloid-DNA mixtures is strong evidence that aspherical polymer shapes 
play a significant role in depletion.  Contrary to previous 
studies,\cite{lekkerkerker-tuinier2011,tuinier-petukhov2002,fleer-tuinier2003}
we conclude that depletion interactions between hard-sphere colloids are not
fully captured by modeling polymers simply as penetrable spheres of an effective size or, 
equivalently, with an effective depletion layer thickness.  Moreover, our approach 
consistently accounts for {\it fluctuations} in polymer shape, in contrast 
to models of spheroidal depletants.\cite{kamien1999,piech-walz2000}

Our approach may be compared with the powerful and elegant method of Bolhuis and Louis 
\etalia\cite{louis2000-prl,bolhuis2001-jcp,bolhuis2001-pre,bolhuis-louis-hansen-prl2002,
bolhuis-louis2002-macromol,louis2002-jcp1,louis2002-jcp2} that models polymers as 
``soft colloids" by replacing a polymer coil with a single particle at the center of mass.
These authors determined the effective pair potential between two polymers and between a 
polymer and a hard sphere by first computing the respective radial distribution 
function $g(r)$ between the centers of mass, via MC simulation of explicit 
segmented polymers  on a lattice, and then inverting $g(r)$ via the Ornstein-Zernike 
integral equation.  From subsequent simulations of a coarse-grained model of 
colloid-polymer mixtures governed by such effective pair potentials, they 
extracted polymer depletion-induced interactions between hard-sphere colloids.
For polymers in a good solvent, whose excluded-volume interactions were modeled 
via self-avoiding walks, comparisons of effective pair potentials derived from
simulations of the explicit and coarse-grained models were in close agreement
for $q\simeq 1$ and in the dilute polymer concentration regime, with deviations 
emerging abruptly at higher concentrations.  Moreover, a computationally practical 
superposition approximation that expresses two-body depletion interactions
in terms of the radially symmetric density profile of a polymer around a 
single colloidal sphere, which for ideal depletants can be implemented as a 
simple convolution integral,\cite{zaccarelli2015} proves nearly as accurate 
as simulations.  Our analysis of polymer shape fluctuations, although here 
limited to ideal polymers, would suggest that the soft-colloid approach succeeds 
largely by capturing, in the effective colloid-polymer potential, an accurate 
representation of the distortion of a polymer coil near a hard, curved surface.

\begin{figure*}
\begin{minipage}{4.1cm}
\includegraphics[width=\columnwidth]{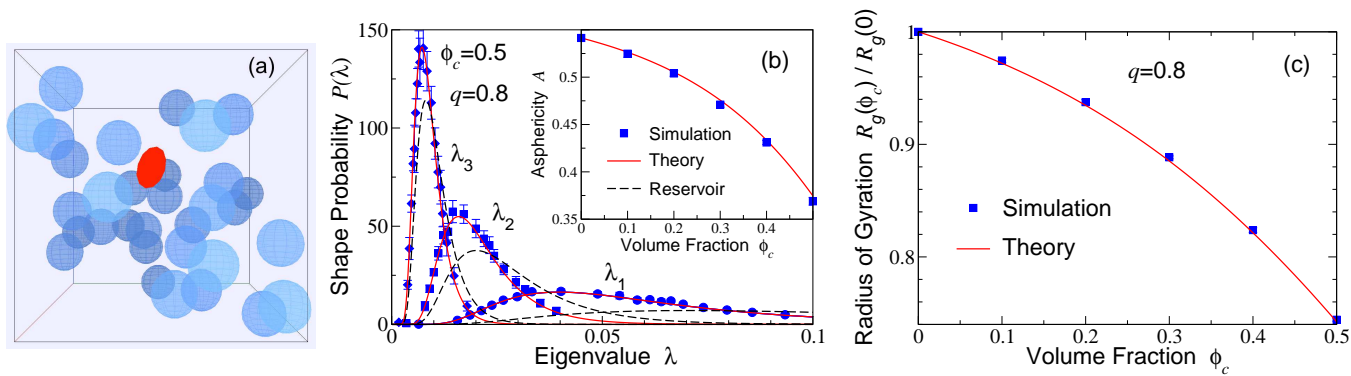}
\end{minipage}
\hspace*{0.01cm}
\begin{minipage}{6.6cm}
\includegraphics[width=\columnwidth]{fig3b.eps}
\end{minipage}
\hspace*{0.2cm}
\begin{minipage}{6.6cm}
\vspace*{-0.1cm}
\includegraphics[width=\columnwidth]{fig3c.eps}
\end{minipage}
\vspace*{-0.2cm}
\caption{
(a) Simulation snapshot depicts colloids (blue spheres) and polymer (red ellipsoid)
in a cubic simulation cell with periodic boundary conditions.
(b) Probability distributions for eigenvalues ($\lambda_1$, $\lambda_2$, $\lambda_3$)
of the gyration tensor of an ideal polymer coil with random-walk segment statistics.  
Simulation data (symbols) 
are compared with predictions of free-volume theory\cite{lim-denton2014} (solid curves)
for an ellipsoidal polymer with uncrowded size ratio $q=0.8$ amidst 216 colloids
of volume fraction $\phi_c=0.5$.  Dashed curves: uncrowded ($\phi_c=0$) distributions
[Eq.~(\ref{Eurich-Maass1})].  Inset: polymer asphericity $A$ vs.~$\phi_c$ [Eq.~(\ref{asphericity})].
(c) Polymer radius of gyration $R_g$ vs.~$\phi_c$. 
}\label{fig-Plambda}
\end{figure*}

Our restriction thus far to the dilute limit, while intended to highlight 
the role of polymer shape fluctuations in depletion interactions, 
raises the important question of how such shape fluctuations may be modified in more 
crowded environments, as in concentrated suspensions and biological cells.  
As a first step toward assessing the impact of crowding on polymer shapes,
we simulated polymers amidst many mobile colloids, now including trial displacements 
of both species. 
Previously, we computed polymer shape distributions, radii of gyration, and asphericities 
in the protein limit ($q\gg 1$), using a coated-ellipsoid approximation for the 
excluded volume.\cite{lim-denton2014}  By applying the exact overlap algorithm, we can now extend 
this analysis to the colloid limit ($q<1$).  Figure~\ref{fig-Plambda} shows results from 
simulations of 216 colloids and one polymer at $q=0.8$ in a cubic box with 
periodic boundary conditions, along with predictions of a free-volume theory 
based on a mean-field approximation for the average volume accessible to an ellipsoid 
in a hard-sphere fluid.\cite{lim-denton2014}
With increasing colloid volume fraction, $\phi_c\equiv (4\pi/3)n_cR_c^3$,
the polymer eigenvalue distributions shift toward contraction 
of the polymer along all three principal axes, while the radius of gyration and 
asphericity decrease, reflecting compactification of the polymer.
These trends imply a decreasing range of pair attraction with increasing colloid concentration.

\section{Conclusions}

In summary, we computed depletion potentials between hard, spherical colloidal particles 
induced by ideal polymers, modeled as fluctuating ellipsoids with random-walk 
segment statistics.  Comparisons with exact theoretical calculations and data from 
both molecular simulations and experiments demonstrate that shape-fluctuating polymers 
induce significantly weaker and longer-ranged interactions than spherical depletants,
even after accounting for particle curvature via an effective depletion layer thickness. 
While the depletion potentials computed here in the dilute limit are not expected to
directly transfer to concentrated colloid-polymer mixtures, in which many-body 
effective interactions may be significant, the ellipsoidal polymer model should be 
applicable at nonzero concentrations.  
When progressively crowded by colloids, polymer coils remain aspherical, 
but become more compact in size and shape.  

Our approach provides a new conceptual framework for interpreting experiments, 
is computationally more efficient than explicit polymer models, and may be adapted 
to model depletion and crowding in mixtures of colloids and excluded-volume 
polymers,\cite{bolhuis-louis2002-macromol,louis2002-jcp1,louis2002-jcp2,
bolhuis-louis-hansen-prl2002, bolhuis2003,doxastakis2005,aarts2002} represented
as self-avoiding random walks\cite{solc1971} in good solvents.
It may be further extended to the protein limit of polymer-nanoparticle mixtures, 
by incorporating an appropriate penetration free energy.\cite{lim-denton2014,lim-denton2015}
Models of shape-fluctuating particles also may be useful for exploring phase behavior 
in polymer nanocomposites and in dispersions of soft colloids, e.g., microgels,
whose shapes deform at high concentrations.\cite{fernandez-nieves2015}

\vspace*{0.5cm}
\noindent{\bf \large Acknowledgments} \\[1ex]
This research was supported by the National Science Foundation (Grant No.~DMR-1106331).

\balance

\bibliographystyle{rsc}

\providecommand*{\mcitethebibliography}{\thebibliography}
\csname @ifundefined\endcsname{endmcitethebibliography}
{\let\endmcitethebibliography\endthebibliography}{}


\end{document}